# Strain Mediated Voltage Control of Magnetic Anisotropy and Magnetization Reversal in Bismuth Substituted Yttrium Iron Garnet Films and Meso-structures

Walid Al Misba [1], Miela Josephine Gross [2,3], Kensuke Hayashi [3,4], Daniel B. Gopman [5], Caroline A. Ross [3], Jayasimha Atulasimha [1,6,7]

[1] Mechanical and Nuclear Engineering, Virginia Commonwealth University, Richmond, VA, USA
[2] Electrical Engineering and Computer Science, Massachusetts Institute of Technology, Cambridge, MA, USA
[3] Department of Materials Science and Engineering, Massachusetts Institute of Technology, Cambridge, MA, USA
[4] Department of Materials Physics, Graduate School of Engineering, Nagoya University, Nagoya, Japan
[5] Materials Science & Engineering Division, National Institute of Standards and Technology, Gaithersburg, MD, USA
[6] Electrical and Computer Engineering, Virginia Commonwealth University, Richmond, VA, USA
[7] Department of Physics, Virginia Commonwealth University, Richmond, VA, USA

**Abstract:**

We report on magnetic anisotropy modulation in Bismuth substituted Yttrium Iron Garnet (Bi-YIG) thin films and mesoscale patterned structures deposited on a PMN-PT substrate with the application of voltage-induced strain. The Bi content is selected for low coercivity and higher magnetostriction than that of YIG, yielding significant changes in the hysteresis loops through the magnetoelastic effect. The piezoelectric substrate is poled along its thickness, which is the [011] direction, by applying a voltage across the PMN-PT/SiO$_2$/Bi-YIG/Pt heterostructure. *In-situ* magneto-optical Kerr effect microscopy (MOKE) shows the modulation of magnetic anisotropy with voltage-induced strain. Furthermore, voltage control of the magnetic domain state of the Bi-YIG film at a fixed magnetic field produces a 90° switching of the magnetization easy axis above a threshold voltage. The magnetoelectric coefficient of the heterostructure is $1.05 \times 10^{-7}$ s m$^{-1}$ which is competitive with that of other ferromagnetic oxide films on ferroelectric substrates such as La$_{0.67}$Sr$_{0.33}$MnO$_3$/PMNPT and YIG/PMN-PZT. Voltage-control of magnetization reversal fields in 5 – 30 μm wide dots and racetracks of Bi-YIG show potential for energy efficient non-volatile memory and neuromorphic computing devices.

## I. Introduction:

Electric field tunability of magnetization is particularly appealing for high density magnetic memory with lower energy consumption [1-3] compared to current-controlled technologies [4,5]. In this regard, multiferroic structures with coupled ferroelectric (FE) and ferromagnetic properties have been examined for their ability to control the electric and magnetic ordering simultaneously through the converse magnetoelectric effect (CME) [6,7]. The required electric current density to write a magnetic random-access memory bit is on the order of $10^{11}$ A/m$^2$ with 10 fJ [8] dissipation compared to 1-100 aJ dissipation in capacitive multiferroic devices [8, 9]. Although single phase multiferroic materials [10,11] are the most direct embodiment of this phenomena, composite heterostructures provide three to four orders of magnitude greater magnetoelectric coupling as well as stability of both polarization and magnetization at room temperature [12,13]. Several mechanisms have been explored for harnessing CME from composite heterostructures, such as transferring mechanical strain from the FE to the ferromagnet [14-16], modulation of the spin-up and spin-down densities of states at the FE-ferromagnet interface [17], and modification of an oxide ferromagnet through voltage-driven oxygen migration [18]. Strain transfer mechanisms demonstrate low heat dissipation per switching cycle and high magnetoelectric coupling coefficients [19].

Relaxor ferroelectric materials such as $(Pb(Mg_{1/3}Nb_{2/3})O_3)_{1-x}-(PbTiO_3)_x$ (PMN-PT) show large piezoelectric coefficients when operated near the morphotropic phase boundary ($x$ = 0.3, for PMN-PT) and have been employed to transfer strain to a ferromagnetic material [20]. Thin films of ferromagnetic materials with low to moderate magnetostriction such as Ni [21], Co [14,15,22,23], CoFeB [16,24], or FeGa [25] have been grown on top of PMN-PT to investigate magnetoelectric effects. The magnetic films in these composites are often amorphous or polycrystalline, enabling electric-field-induced magnetoelastic anisotropy to exceed magnetocrystalline anisotropy for 90° rotation of the magnetic easy axis. Complete 180° switching was demonstrated in patterned Co/PMN-PT by sequentially applying voltages in the electrode pairs [26].

In contrast to ferromagnets, many ferrimagnetic oxides offer more efficient and faster control of magnetization due to their low damping and moderate saturation magnetization. Ferrimagnetic oxides such as yttrium iron garnet (YIG) and rare-earth iron garnets (REIGs) have been used to demonstrate spin wave propagation and spin torque phenomena [27,28]. In addition, the saturation magnetization, magnetostriction, anisotropy and Gilbert damping parameter can be modified by inserting rare earth ions [29]. Despite these advantages, the growth of ferrimagnetic garnets on piezoelectric compounds poses a significant challenge due to lattice incompatibility, thus limiting the potential to harness the benefit of electrical control. We developed a thin film processing strategy centered around the growth of a thin, amorphous $SiO_2$ spacer that enables growth of high quality polycrystalline ferrimagnetic garnets on bulk piezoelectric substrates [30]. This materials processing advance allowed us to evaluate strain-induced anisotropy modulation of yttrium-substituted dysprosium iron garnet (Y-DyIG) film crystallized on PMN-PT [28], when the PMN-PT was poled.

Bi-YIG has certain advantages over other REIGs due to its low loss tangent [31], large domain wall velocities [32], low Gilbert damping [33,34] and magneto-optical activity [35] making it important to many applications in magneto-optics such as optical isolators [36], electrical current sensors [37], spintronics [38], and magnonic devices such as spin wave carriers [39]. In this study, we demonstrate magnetoelectric control of Bi-YIG thin films and patterned microstructures on PMN-PT. *In-situ* magneto-optical Kerr microscopy (MOKE) studies show electrical-field control of magnetization and electrically tunable magnetic properties. We show that the magnetic easy axis of Bi-YIG films can be reoriented by 90° degrees under an applied electric field. Furthermore, we show voltage control of magnetization reversal of micron-sized magnetic dots and racetracks. These results suggest a role for Bi-YIG in energy efficient voltage-controlled memory and neuromorphic devices [40-42].

**II. Sample Growth and Characterization Methods**

Following process conditions used for DyIG [28], a series of 0.5 mm thick, (011)-oriented PMN-PT $[(PbMg_{0.33}Nb_{0.67}O_3)_{1-x}(PbTiO_3)_x$; $x$=0.29-0.33] substrates were coated with an amorphous, 2.4 nm thick $SiO_2$ buffer layer by radio frequency magnetron sputtering. The 45.6 nm thick Bi-YIG films were grown on the PMN-PT/$SiO_2$ heterostructures in addition to fused silica ($SiO_2$) and (100)-oriented Si substrates using pulsed laser deposition (PLD) at room temperature by co-deposition from stoichiometric YIG ($Y_3Fe_5O_{12}$) and BFO ($BiFeO_3$) targets to yield a composition of $Bi_{2.13}Y_{1.40}Fe_5O_x$ [43]. This high-Bi composition has a higher magnetostriction which counteracts the in-plane shape anisotropy in tensile-strained films and can even lead to out-of-plane easy axis in Bi-YIG on fused silica [43]. A 248 nm KrF excimer laser was used at an energy of 600 mJ, a repetition rate of 10 Hz, and was focused to a fluence of about 2 J cm$^{-2}$ at each target. The laser shots on each target were adjusted based on the calibrated growth rates. The chamber was pumped to a base pressure of 1.33 mPa (1 x 10$^{-5}$ Torr) and an oxygen pressure of 2.7 Pa (20 mTorr) was maintained during the deposition. The films then underwent an *ex-situ* anneal in a furnace for 72 h at 600 °C in to crystallize the garnet.

Grazing incidence x-ray diffraction (GIXD) and film thickness x-ray reflectivity (XRR) measurements were performed using a Rigaku Smartlab Multipurpose Diffractometer with a Cu Kα x-ray source [40]. Magnetic hysteresis curves of the films were measured using a Digital Measurements Systems Vibrating Sample Magnetometer Model 1660 with field applied both within the plane and normal to the plane of the substrate. A Zeiss Merlin high-resolution scanning electron microscope (SEM) was used to examine the grain structure. Films were patterned into ellipses and tracks with minimum dimension of 5 μm by photolithography using a direct-write Heidelberg uMLA exposure system and dry etching using an ion beam etch system.

Figure 1 summarizes the structural and magnetic properties of the Bi-YIG film grown on $SiO_2$-buffered PMN-PT. GIXD is sensitive to the diffraction peaks from the film and shows a set of peaks characteristic of polycrystalline garnet, Fig. 1a, without a strong texture. Consistent with previous observations [33,43,44], the film possesses a preferred magnetization direction within the plane, a saturation magnetization of 101 ±5 kA/m and an in-plane coercivity of 10 ±5 mT. SEM micrographs are presented for the Bi-YIG deposited on Si and PMN-PT/ $SiO_2$ in Fig. 2. A low area fraction of small amorphous regions is observed within the predominantly crystallized specimen over the extent of the examined region. Grain sizes are mainly in the range of 1-3 μm and some show a radiating pattern characteristic of low angle grain boundaries that develop as the grains grow [45]. Crystallization of Bi-YIG over an amorphous buffer layer avoids epitaxy with the PMN-PT, which would yield an orthoferrite-structured film instead of garnet. This buffer layer technique can be extended to other garnet compositions and substrate materials, enabling integration with, for example, amorphous dielectric layers within semiconductor chips.

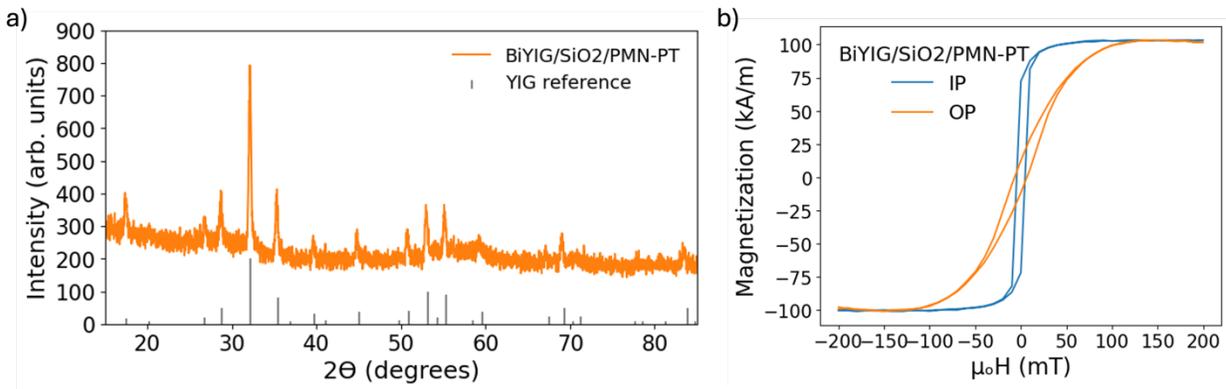

**Fig. 1: a.** GIXD diffraction image shows Bi-YIG growth on $SiO_2$/PMN-PT substrate. Data has been shifted vertically for clarity. Reference powder diffraction peaks for YIG are indicated. **b.** Hysteresis loops taken via vibrating sample

magnetometry of the BiYIG/SiO$_2$/PMN-PT sample. The curves were measured with the field applied out of plane (OP) and in plane (IP) to the sample surface.

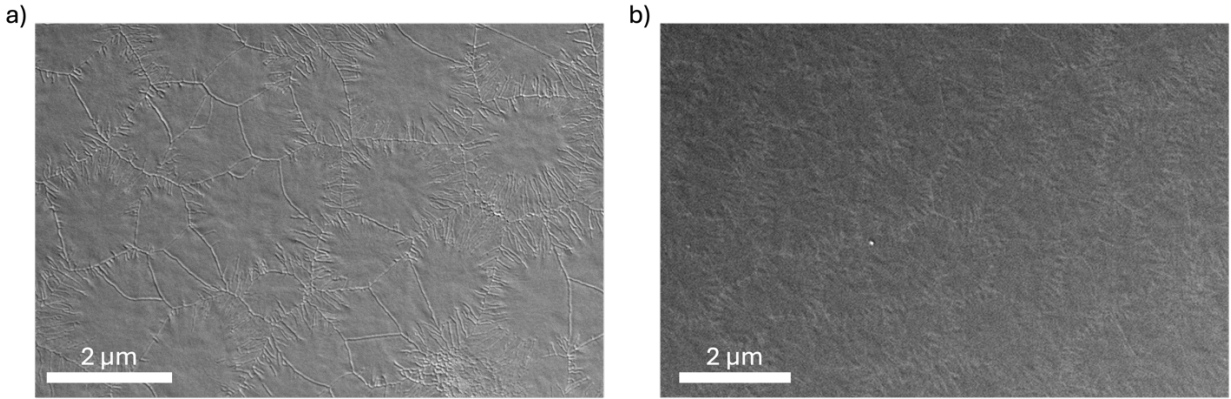

**Fig. 2:** Top surface SEM images of **a.** Si/Bi-YIG and **b.** PMN-PT/ SiO$_2$/Bi-YIG.

### III. Magnetic Hysteresis Modulation with Strain

The magnetic properties of the ferromagnetic material in a FE-ferromagnet heterostructure can be modulated by utilizing the piezoelectric properties of the FE crystal. Applying a voltage across the thickness of the PMN-PT (i.e. along the film normal, defined as $\hat{z}$, the [011] direction) generates an electric field $E$ leading to a piezoelectric strain of different signs in the FE-crystal along the two orthogonal in-plane directions, $\hat{x}$, [100] and $\hat{y}$, the [01$\bar{1}$] direction as shown in the heterostructure schematic in Fig. 3a. When the electric field is zero, the Bi-YIG shows isotropic magnetic behavior within the plane. An electric field along $\hat{z}$ leads to compressive strain along $\hat{x}$ and tensile strain along $\hat{y}$ of the substrate, breaking the degeneracy of the Bi-YIG in-plane hysteresis loops.

To characterize the magnetoelectric behavior of the composite, we first poled the PMN-PT/SiO$_2$/Bi-YIG by applying 450 V along $\hat{z}$ ($E$ = 0.9 MV/m) for 90 min, and then set the voltage to zero. We then applied voltages in 50 V increments, capturing in-plane hysteresis loops using the longitudinal MOKE magnetometry signal at each voltage, as shown in Fig. 3a and Fig. 3b. Blue light (wavelength ≈ 465 nm) was used because Bi-YIG has a high MOKE response at this wavelength. The as-deposited sample is isotropic in plane and showed similar magnetic hysteresis loops along $\hat{x}$ and $\hat{y}$. Poling and subsequent relaxation leads to a remanent strain in the PMN-PT, which is tensile along $\hat{x}$ and compressive along $\hat{y}$ [30,46]. Hence, after poling the Bi-YIG shows a harder magnetization direction (lower remanence and squareness) along $\hat{x}$ and an easier direction along $\hat{y}$ at 0 V compared with the as-grown state, consistent with a negative magnetostriction [47,48]. When voltage is subsequently applied to the poled sample, the PMN-PT experiences increasing compressive strain along $\hat{x}$ due to the negative piezoelectric coefficient, $d_{31}$ of PMN-PT [46], and tensile strain along $\hat{y}$ due to the positive piezoelectric coefficient $d_{32}$ of PMN-PT. This leads to an anisotropy reorientation in the Bi-YIG with $\hat{x}$ becoming the easy in-plane direction for a sufficiently large voltage.

With respect to $\theta$ and $\varphi$, the polar and azimuthal angles of magnetization shown in Fig. 3a, the magnetoelastic energy can be expressed as $F_{me} = -\frac{3}{2}\lambda_s \frac{Y}{1+\vartheta}\varepsilon_{xx}\sin^2\theta\cos^2\varphi - \frac{3}{2}\lambda_s \frac{Y}{1+\vartheta}\varepsilon_{yy}\sin^2\theta\sin^2\varphi$,

where $\varepsilon_{xx}$ and $\varepsilon_{yy}$ are the strains along $\hat{x}$ and $\hat{y}$, $Y$ is the Young's modulus, and $\vartheta$ is the Poisson's ratio of Bi-YIG. There is no stress along $\hat{z}$ due to the free boundary condition at the top surface. The saturation magnetostriction coefficient of polycrystalline Bi-YIG is negative, $\lambda_s \approx -4\times 10^{-6}$ [47,48], from which it follows that the magnetoelastic free energy is reduced when the magnetization is aligned along a compressively strained direction.

Fig. 3a shows the hysteresis loops become increasingly square along $\hat{x}$ as the voltage is increased from 0 V to 450 V, an indication of the development of a magnetic easy axis along that direction. The coercive field increases from $25\pm 2$ mT at 0 V to $27\pm 2$ mT at 450 V, and the saturation field decreases from $77\pm 2$ mT at 0 V to $57\pm 2$ mT at 450 V. Opposite trends are observed along $\hat{y}$: the loop becomes less square with increasing voltage and the coercivity decreases from 30 mT $\pm$ 2 mT at 0 V to 24 mT$\pm$ 2 mT at 450 V (Fig. 3b). A high squareness ratio, defined as $Sq = M_r/M_s$ where $M_r$ and $M_s$ are the remanent and saturation magnetization respectively, indicates the easy axis. Sq increases (decreases) along $\hat{x}$ ($\hat{y}$) with increasing voltage (Fig. 3a-b). A butterfly-like hysteresis loop is observed for $M_r$ vs. $V$ (Fig. 3c) illustrates the magnetoelectric coupling between the applied voltage and remanent magnetization. The loop measured along the poling direction, $\hat{z}$, using polar MOKE shows little change with voltage (Fig. 3d).

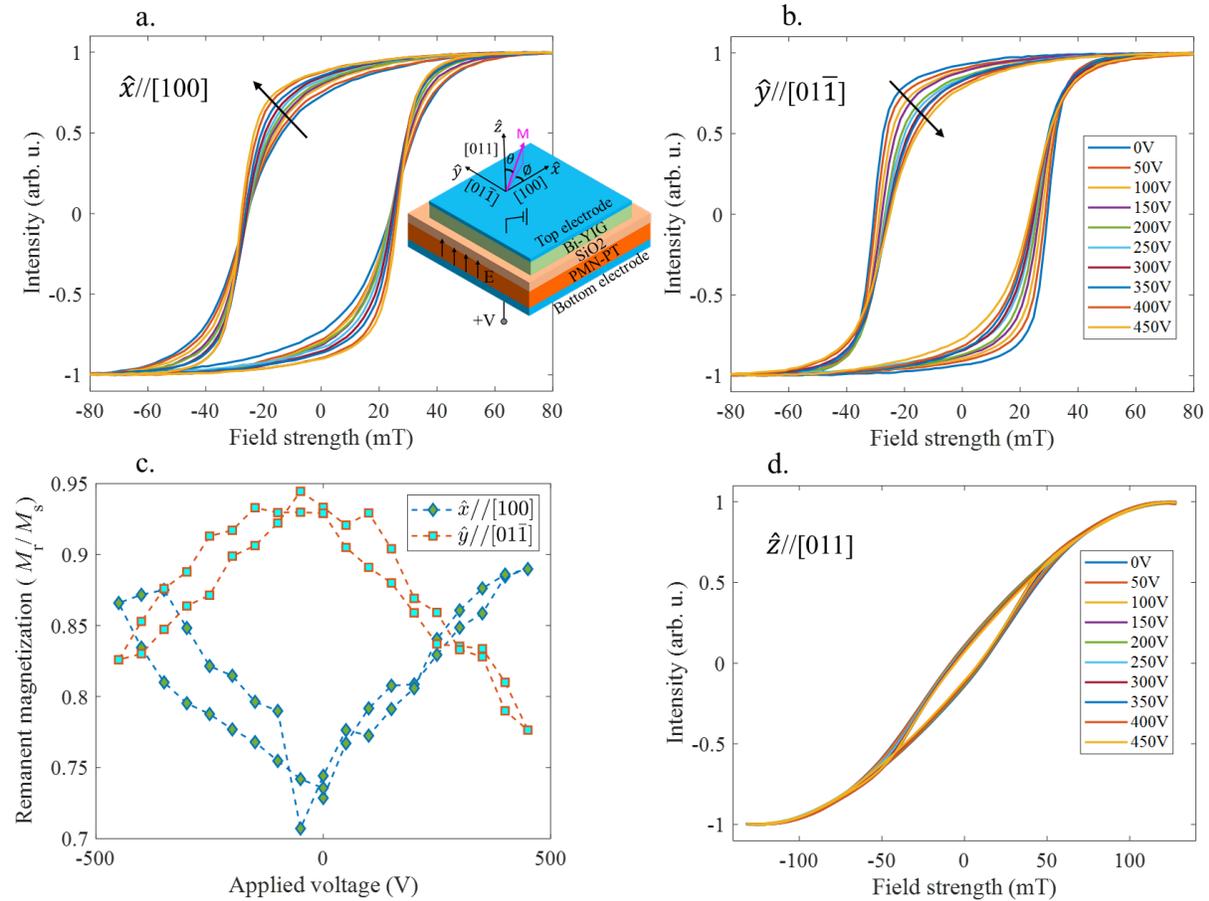

**Fig.3: a, b.** Hysteresis loops obtained from MOKE magnetometry for different voltages applied along the thickness of the heterostructure, PMN-PT/SiO$_2$/Bi-YIG when the magnetic field is applied along the in-plane direction **a.** $\hat{x}$. **b.** $\hat{y}$. Black arrows indicate the trend for increasing voltage. The inset in (a) shows a schematic of the heterostructure with the direction of the applied voltage, principal axes and the polar angle, $\theta$ and azimuthal angle, $\varphi$ of the BiYIG

film magnetization, M.  **c**. Ratio of remanent and saturation magnetization vs the applied voltage for both in-plane directions, $\hat{x}$ and $\hat{y}$.  **d**. Hysteresis loops as a function of voltage obtained from polar MOKE for the out of plane direction, $\hat{z}$.

In Fig. 4 we analyze the magnetization switching of the poled heterostructure for two cases, 0 V and 450 V for the in-plane directions $\hat{x}$ and $\hat{y}$. Initially, a reference background image is taken from which the images acquired at different magnetic fields are subtracted. At positive saturation, +88 mT, predominantly white contrast domains are observed. The images in Fig. 4b and 4d correspond to the positions marked on the hysteresis loops in Fig. 4a and 4c respectively.

As the external field is increased in the negative direction, reversal is indicated by the black contrast. For fields applied along $\hat{x}$, when the voltage is 0 V, the switching corresponds to a gradual change in contrast, and no significant domain wall propagation is observed along the hard $\hat{x}$ axis (compare the images at -29.40 mT and -33.40 mT in bottom panels of Fig. 4b which corresponds to an easy $\hat{x}$ and therefore a sharper transition with more prominent domain wall nucleation and propagation features). Along the $\hat{y}$ direction the easy axis switching process occurs with domain wall nucleation and propagation at 0 V, but a gradual contrast change is observed at 450 V, consistent with $\hat{y}$ becoming a hard axis with increasing voltage.

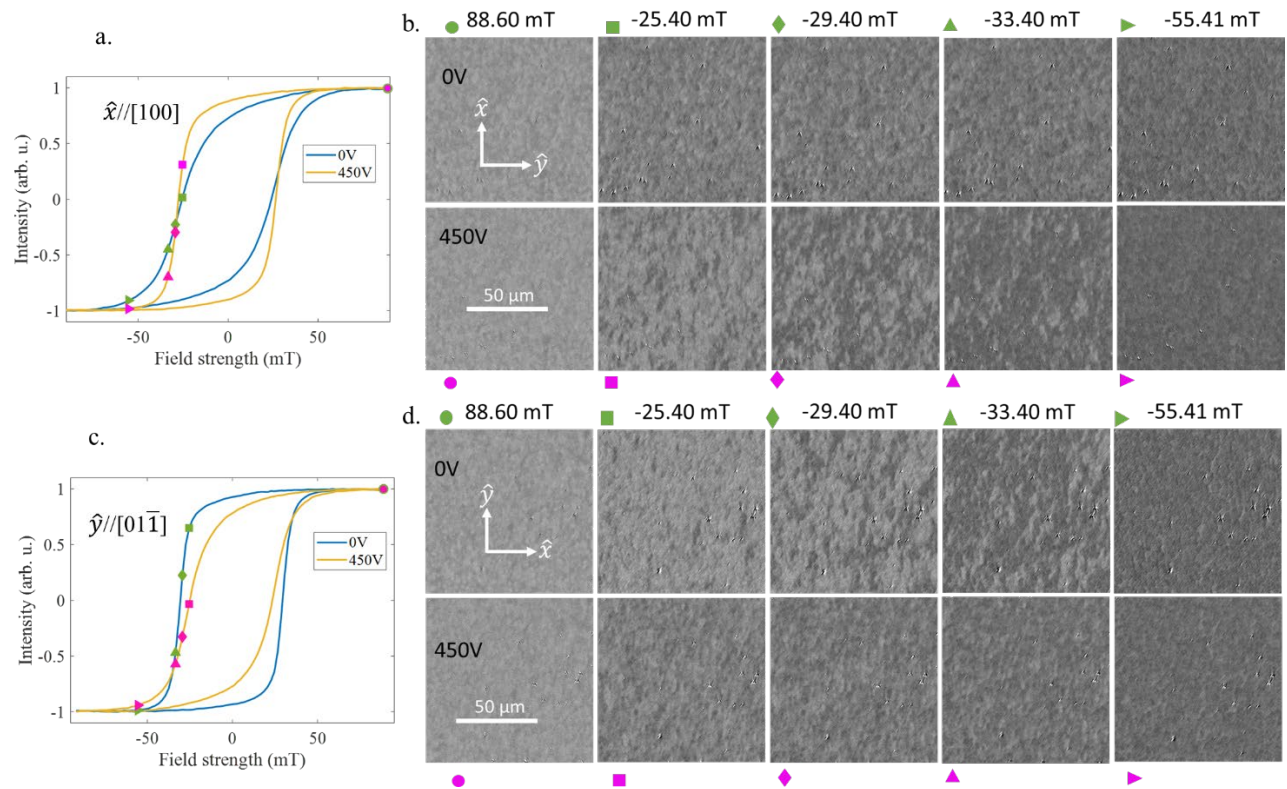

**Fig. 4: a**. Hysteresis curves with external fields applied along the in-plane direction $\hat{x}$ when the heterostructure is subjected to an applied voltage of 0 V and 450 V. **b**. longitudinal MOKE images showing magnetization reversal process. The corresponding field values for which the images are taken are also marked in the hysteresis loops using green and purple polygon markers. **c**. hysteresis loops for in-plane direction $\hat{y}$ for 0 V and 450 V and **d.** corresponding magnetization reversal images.

## IV. Magnetization Reversal with Electric Field

Fig. 5 shows voltage control of magnetic domains at a fixed magnetic field. The sample was first saturated by applying a field of -70 mT, then the field was set to +27 mT and the corresponding domain patterns were observed as a function of voltage. The field of +27 mT was selected because it is close to the coercive field for both of the in-plane directions and led to significant changes in the domain pattern with voltage. To cycle the electric field, the sample was first poled by applying 450 V and subsequently relaxed to 0 V. For image acquisition in the $\hat{x}$-direction, 450 V is applied, and followed by a magnetic field of -70 mT while the voltage is maintained at 450 V. In this configuration, domains with black contrast are predominant. The external field is then increased to +27 mT. White-contrast domains indicate the onset of reversal, which increases as the voltage is reduced stepwise in increments of 50 V, from 450 V to 0 V, while keeping the magnetic field constant at 27 mT. The domain pattern shows little change for voltages below 100 V. The behavior is explained by the $\hat{x}$ axis being the easy axis at 450 V but becoming less easy as the voltage decreases until it becomes the hard axis by 0 V. The reduction in anisotropy along the $\hat{x}$ axis leads to magnetization rotation towards the $\hat{y}$-axis, reduction in domain sizes and a weakening of contrast. At low voltages, $\hat{x}$ becomes a magnetically hard direction and the magnetization orientation within the domains is governed by the balance between the magnetoelastic anisotropy and the Zeeman energy. An analogous but opposite trend is found when the field is applied along $\hat{y}$, shown in Fig. S1 of the Supplementary Information. Thus, in Fig 5, the preferred axis of magnetization shifts from $\hat{y}$ to $\hat{x}$ as the voltage is increased from 0 V to 450 V and a 90° switching of the easy axis is accomplished.

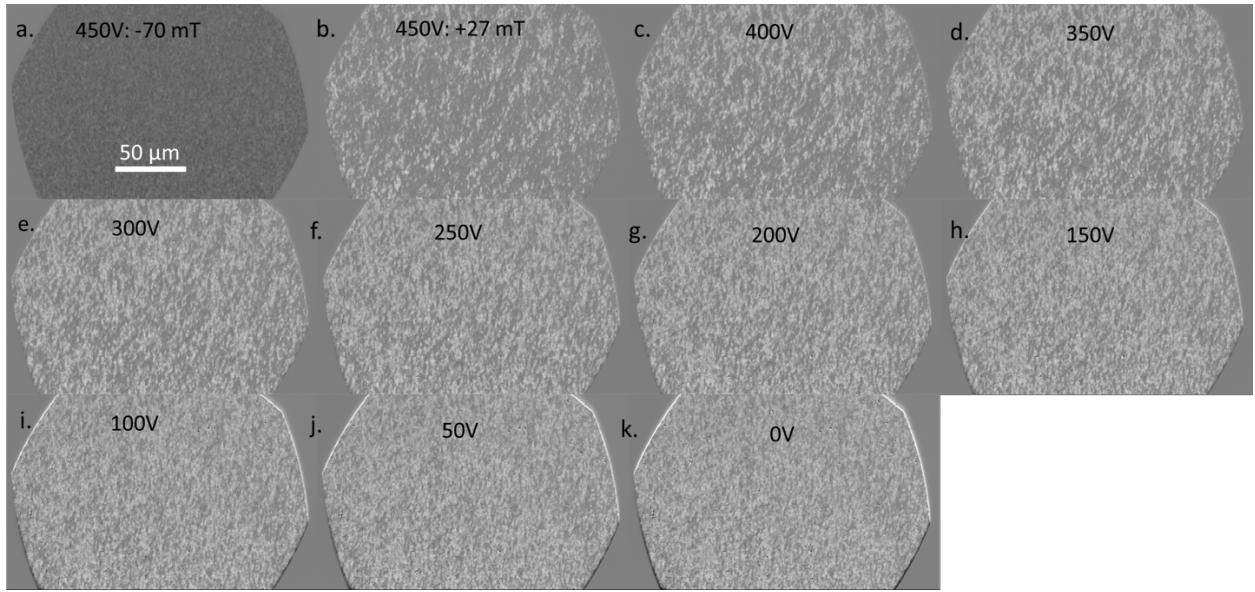

**Fig. 5:** MOKE images showing the saturated domains and reversal of the domains for varying amplitude voltages at a fixed reversal field along the in-plane direction $\hat{x}$ // [100]. a. The sample is poled at 450 V and saturated with -70 mT field. The external field is then fixed at + 27 mT, while the voltage remains at b. 450 V and decreases to c. 400 V d. 350 V e. 300 V f. 250 V g. 200 V h. 150 V i. 100 V j. 50 V and k. 0 V.

The magnetoelectric coefficient, $\alpha_E = \mu_0 \frac{\Delta M}{\Delta E}$ [49] of the PMN-PT/SiO$_2$/Bi-YIG is calculated from Fig. 3c (in-plane direction, $\hat{x}$), under the condition of zero applied field. Thus, $\Delta M$ is the change in the remanent magnetization, $\Delta M_r$ and, $\Delta E = \frac{\Delta V}{t}$, where $t$ is the thickness of the heterostructure, $E$ is the electric field and $V$ is the applied voltage across the heterostructure. The highest value of $\alpha_E$ is determined to be 1.05 $\times 10^{-7}$ s m$^{-1}$, which is achieved when the voltage is changed from 0 V to -50 V with a maximum change in remanent magnetization of $\approx$ 8.36 kA/m. Table I compares the ME coefficient of PMN-PT/SiO$_2$/Bi-YIG (this work) with those of other ferroelectric/magnetic bilayers. Table I shows that the estimated magnetoelectric coefficient of PMN-PT/SiO$_2$/Bi-YIG is comparable with previously examined ferroelectric/ magnetic oxide systems, though lower than ferroelectric/magnetic metal systems.

To compare the dynamical properties of polycrystalline Bi-YIG with that measured on single crystal films we measured the ferromagnetic resonance of the fused silica/ Bi-YIG sample at frequencies between 5 GHz and 9 GHz (Supplementary Figure S2). The frequency range is too small to determine damping, but the linewidths were approximately 200 mT compared with as low as 5 mT for single crystal epitaxial Bi-YIG grown on a garnet substrate [33,34]. The higher linewidth is attributed to the anisotropy distribution arising from the polycrystalline microstructure, the residual amorphous non-magnetic regions and the high film stress [43] and can likely be reduced by optimizing the process or composition.

Table I: comparison of magnetoelectric coefficient

| Ferroelectric/Magnetic Materials | Magnetic Film Type | Magnetoelectric Coefficient (s m$^{-1}$) |
|---|---|---|
| PMN-PT/ Bi-YIG | oxide | $1.05 \times 10^{-7}$ [this work] |
| PMN-PT/ Y-DyIG | oxide | $2.8 \times 10^{-7}$ [28] |
| PMN-PT/ YIG | oxide | $5.4 \times 10^{-9}$ [50] |
| PMN-PZT/ YIG | oxide | $1.8 \times 10^{-7}$ [51] |
| PMN-PT/ La$_{0.7}$Sr$_{0.3}$MnO$_3$ | oxide | $6.4 \times 10^{-8}$ [52] |
| PMN-PT/ FeGa | metallic | $2.7 \times 10^{-6}$ [53] |
| PMN-PT/ Co$_2$FeSi | metallic | $1 \times 10^{-5}$ [54] |
| BaTiO$_3$/ FeRh | metallic | $1.6 \times 10^{-5}$ [55] |

Finally, to check the reproducibility of the strain-dependent hysteresis of the PMN-PT/SiO$_2$/Bi-YIG, we investigated another sample with 55 nm Bi-YIG thickness. This sample shows a similar trend in strain induced anisotropy modulation and a magnetoelectric coefficient of $0.9 \times 10^{-7}$ s m$^{-1}$.

**V. Voltage Control of Magnetism in Bi-YIG Microstructures:**

The effect of electric field is studied on PMN-PT/SiO$_2$/Bi-YIG microstructures fabricated using photolithography and dry etching. Patterned ellipses and racetracks were examined to see the effect of the magnetoelectric coupling on voltage control of magnetization switching and domain evolution. Fig. 6 presents the magnetization switching in ellipses when different voltages are applied across the microstructures. The easy axes of the ellipses are parallel to the in-plane substrate direction $\hat{x}$. The samples are first poled at 450 V and then relaxed. The ellipses are saturated with -20 mT field applied along $\hat{x}$ and the field is increased to 20 mT in 0.5 mT intervals to observe the magnetization evolution during switching. The magnetization in the ellipses is switched at lower field along the $\hat{x}$ direction when the voltage is modified from 450 V to 0 V. For example, the ellipses are mostly switched from negative $\hat{x}$ direction (black contrast) to positive $\hat{x}$ direction (white contrast) at 8 mT field for 0 V. Thus, the switching field of the patterned magnets can be controlled by modifying the electric field across the PMN-PT substrate similarly to the Bi-YIG films. At high voltage ($V$ = 450 V), the Bi-YIG films exhibited higher coercivity along the $\hat{x}$ direction which, in turn, decreases as the voltage is reduced.

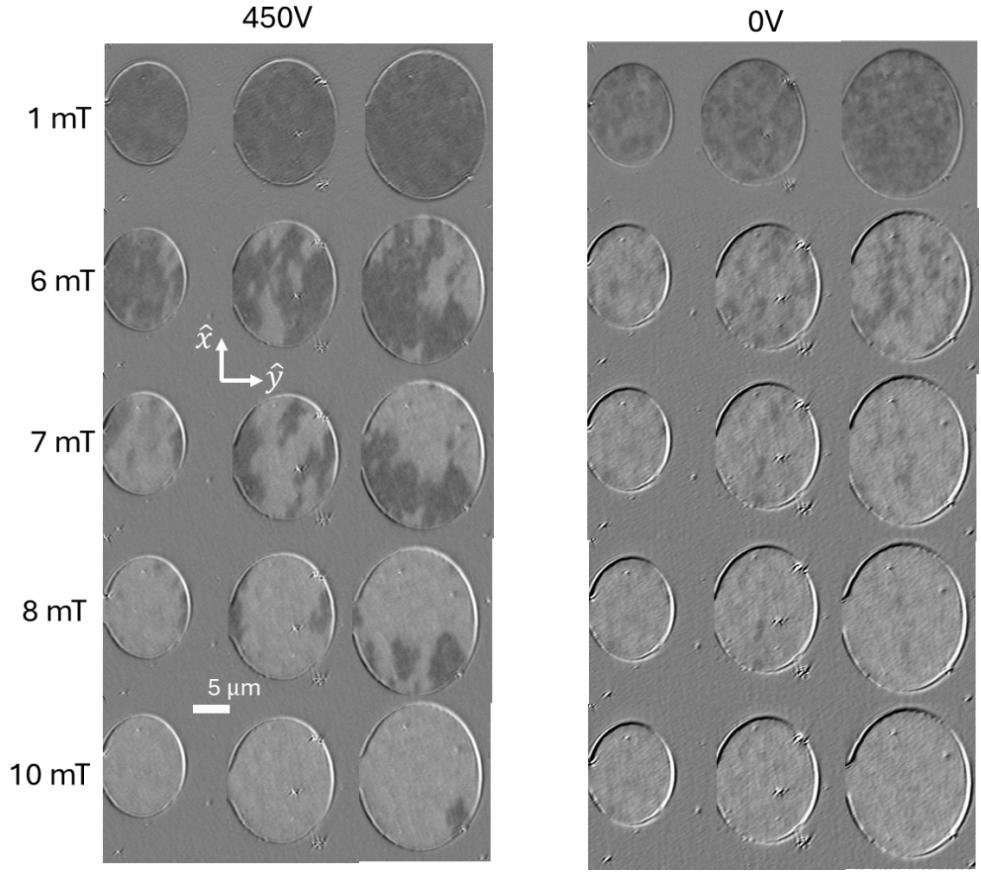

**Fig. 6:** Snapshots of magnetization switching in elliptical microscale magnets of Bi-YIG patterned on PMN-PT for two different voltages. The magnetic fields are applied along in-plane direction $\hat{x}$ which is parallel to the easy axes of the ellipses. The elliptical magnets switched at lower field (8 mT) for $V$ = 0 V compared to $V$ = 450 V.

Next, the effect of electric field was studied in racetracks with their long axis parallel to the $\hat{x}$ direction. Domain wall nucleation and propagation are observed due to the field applied along the $\hat{y}$ direction. Both the racetrack and the nucleation pads were saturated to -30 mT and the field was increased to + 30 mT in 1 mT increments. The corresponding MOKE images are presented in Fig. 7 for the poled sample under two different voltages, $V$ = 0 V and $V$ = 450 V. With the increase in voltage, the nucleation field of the domain wall decreases. Stronger movement of domain walls from the pad to racetrack is observed at 7 mT for 0 V. However, the domain wall propagation becomes easier at 450 V and starts to propagate with as little as 6 mT field.

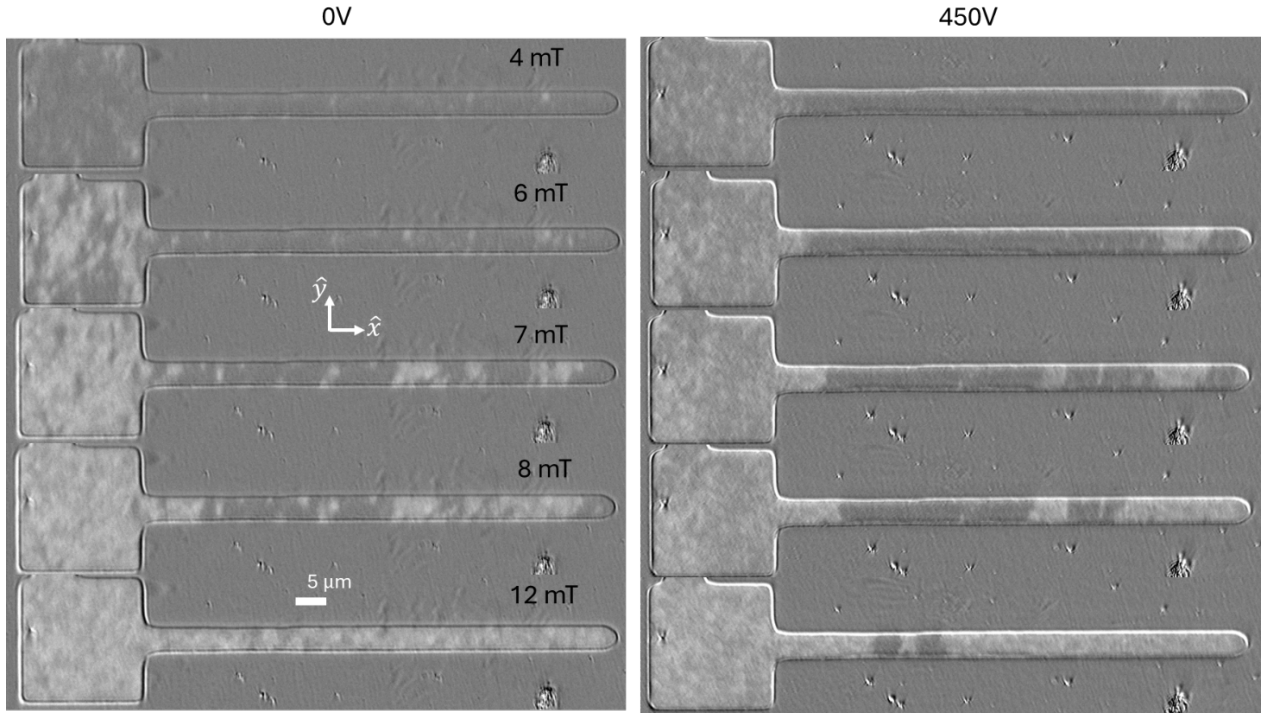

**Fig. 7:** Snapshots of domain wall nucleation and propagation in 5 µm wide racetracks of Bi-YIG patterned on PMN-PT for two different voltages. The magnetic fields are applied along in-plane direction $\hat{y}$. The domain wall nucleated in the pads propagate along the racetracks at 6 mT when it is subjected to a voltage of 450 V.

## VI. Summary and Conclusion

In summary, we have shown 90 ° switching of the magnetization easy axis of a multiferroic heterostructure, PMN-PT/SiO$_2$/Bi-YIG by using voltage-induced strain. The Bi-YIG film was fabricated using pulsed layer deposition and the ratio of Bi and YIG was selected for a high Bi content. An intermediate SiO$_2$ buffer layer is deposited between PMN-PT and Bi-YIG to avoid growth of perovskite phases and thereby facilitate garnet crystallization. MOKE magnetometry shows domain wall nucleation and propagation in Bi-YIG films with the application of electric fields and strain mediated electric voltage control of magnetization reversal fields in patterned mesostructures. Although the magnetoelectric coefficient is moderate compared to heterostructures combining magnetostrictive metals with PMN-PT, it compares well with other oxide ferrimagnet/PMN-PT structures. The magnetoelectric response can stimulate novel devices that use resonant effects [56-58], and lead to energy efficient magnetic memory and neuromorphic computing devices [59,60].

**Acknowledgement:**

WAM and JA acknowledge support from NSF ECCS grant #1954589, ECCS-EPSRC grant #2152601, CISE: SHF: Small grant #1815033 and NSF MRI grant #2117646, and the use of Virginia microelectronic center (VMC) and VCU nano characterization center (NCC). CAR and MJG acknowledge support from NSF award ECCS 2152528 and the use of shared facilities of MIT.nano.

**References:**

[1] F. Matsukura, Y. Tokura, H. Ohno, Control of Magnetism by Electric Fields. Nat. Nanotechnol., vol. 10, 209–220 (2015).

[2] S. Bandyopadhyay, J. Atulasimha, Nanomagnetic and Spintronic Devices for Energy-Efficient Memory and Computing, 1st ed.; Wiley, 2016.

[3] S. Bandyopadhyay, J. Atulasimha, A. Barman, Magnetic Straintronics: Manipulating the Magnetization of Magnetostrictive Nanomagnets with Strain for Energy-Efficient Applications, *Applied Physics Reviews*, vol. 8, 041323 (2021).

[4] J. C. Slonczewski, Current-driven excitation of magnetic multilayers, J. Magn. Magn. Mater., vol. 159, L1 (1996).

[5] H. Kubota, A. Fukushima, K. Yakushiji, T. Nagahama, S. Yuasa, K. Ando, H. Maehara, Y. Nagamine, K. Tsunekawa, D. D. Djayaprawira, N. Watanabe, and Y. Suzuki, Quantitative measurement of voltage dependence of spin-transfer torque in MgO, Nature Physics, vol. 4, pp. 37–41 (2008).

[6] N. A. Spaldin and M. Fiebig, The Renaissance of Magnetoelectric Multiferroics, vol. 309, no. 5733, pp. 391-392 (2005)

[7] R. Ramesh and N. A. Spaldin, Multiferroics: progress and prospects in thin films, Nature Materials, vol. 6, pp. 21–29 (2007)

[8] N. A. Spaldin, and R. Ramesh, Advances in magnetoelectric multiferroics, Nature Materials, vol. 18, 203–212 (2019)

[9] J. Atulasimha, S. Bandyopadhyay, Bennett clocking of nanomagnetic logic using multiferroic single-domain nanomagnets, Appl. Phys. Lett., vol. 97, 173105 (2010).

[10] N. Hur, S. Park, P. A. Sharma, J. S. Ahn, S. Guha, S.-W. Cheong, Electric Polarization reversal and Memory in a Multiferroic Material Induced by Magnetic Fields. Nature, vol. 429, 392–395 (2004)

[11] V. J. Folen, G. T. Rado, E. W. Stalder, Anisotropy of the Magnetoelectric Effect in Cr2O3. Phys. Rev. Lett., vol. 6, 607–608 (1961).

[12] W. Eerenstein, M. Wiora, J. L. Prieto, J. F. Scott, N. D. Mathur, Giant Sharp and Persistent Converse Magnetoelectric Effects in Multiferroic Epitaxial Heterostructures. Nat. Mater., vol. 6, 348– 351 (2007).

[13] J. T. Heron, J. L. Bosse, Q. He, Y. Gao, M. Trassin, L. Ye, J. D. Clarkson, C. Wang, J. Liu, S. Salahuddin, D. C. Ralph, D. G. Schlom, J. Iñiguez, B. D. Huey, R. Ramesh, Deterministic Switching of Ferromagnetism at Room Temperature Using an Electric Field. Nature, vol. 516, pp. 370–373 (2014)

[14] N. D'Souza, M. S. Fashami, S. Bandyopadhyay, J. Atulasimha, Experimental Clocking of Nanomagnets with Strain for Ultra Low Power Boolean Logic, Nano Letters, vol. 16, pp. 1069–1075 (2016)

[15] V. Sampath, N. D'Souza, D. Bhattacharya, G. M. Atkinson, S. Bandyopadhyay, and J. Atulasimha, Acoustic wave-induced magnetization switching of magnetostrictive nanomagnets from single-domain to nonvolatile vortex states, Nano Lett., vol. 16, 5681 (2016).

[16] S. Zhang, Y. G. Zhao, P. S. Li, J. J. Yang, S. Rizwan, J. X. Zhang, J. Seidel, T. L. Qu, Y. J. Yang, Z. L. Luo, Q. He, T. Zou, Q. P. Chen, J. W. Wang, L. F. Yang, Y. Sun, Y. Z. Wu, X. Xiao, X. F. Jin, J. Huang, C. Gao, X. F. Han, R. Ramesh, Electric-Field Control of Nonvolatile Magnetization in Co40Fe40B20/Pb(Mg1/3Nb2/3)0.7Ti0.3O3 Structure at Room Temperature. Phys. Rev. Lett., vol. 108, 137203 (2012)

[17]. C.-G. Duan, J. P. Velev, R. F. Sabirianov, Z. Zhu, J. Chu, S. S. Jaswal, E. Y. Tsymbal, Surface Magnetoelectric Effect in Ferromagnetic Metal Films. Phys. Rev. Lett., vol. 101, 137201 (2008).

[18] U. Bauer, L. Yao, A. J. Tan, P. Agrawal, S. Emori, H. L. Tuller, S. van Dijken, G. S. D. Beach, Magneto-ionic Control of Interfacial Magnetism. Nat. Mater., vol. 14, 174–181 (2015)

[19] J.-M. Hu, C.-G. Duan, C.-W. Nan, L.-Q. Chen, Understanding and Designing Magnetoelectric Heterostructures Guided by Computation: Progresses, Remaining Questions, and Perspectives. npj Comput. Mater., vol. 3, 18 (2017)

[20] S. Zhang, F. Li, High performance ferroelectric relaxor-PbTiO3 single crystals: Status and perspective. J. Appl. Phys., vol. 111, 031301 (2012)

[21]. S. Lindemann, J. Irwin, G.-Y. Kim, B. Wang, K. Eom, J. Wang, J. Hu, L.-Q. Chen, S.-Y. Choi, C.-B. Eom, M. S. Rzchowski, Low-voltage magnetoelectric coupling in membrane heterostructures, Sci. Adv., vol. 7, eabh2294 (2021)

[22] D. B. Gopman, P. Chen, J. W. Lau, A. C. Chavez, G. P. Carman, P. Finkel, M. Staruch, R. D. Shull, Large Interfacial Magnetostriction in (Co/Ni)4/Pb(Mg1/3Nb2/3)O3–PbTiO3 Multiferroic Heterostructures, ACS Applied Materials & Interfaces, vol. 10, no. 29 (2018)

[23] Y. Hsiao, D. B. Gopman, K. Mohanchandra, P. Shirazi and C. S. Lynch, Effect of interfacial and edge roughness on magnetoelectric control of Co/Ni microdisks on PMN-PT(011), Scientific Reports, vol. 12, Art. no. 3919 (2022)

[24] Z. Zhao, M. Jamali, N. D'Souza, D. Zhang, S. Bandyopadhyay, J. Atulasimha, and J.-P. Wang, Giant voltage manipulation of MgO-based magnetic tunnel junctions via localized anisotropic strain: A potential pathway to ultra-energy-efficient memory technology, Appl. Phys. Lett., vol. 109, 092403 (2016)

[25] A. Begué´ and M. Ciria, Strain-Mediated Giant Magnetoelectric Coupling in a Crystalline Multiferroic Heterostructure, ACS Appl. Mater. Interfaces, vol. 13, 6778–6784 (2021)

[26] A. Biswas, H. Ahmad, J. Atulasimha, and S. Bandyopadhyay, Experimental demonstration of complete 180º reversal of magnetization in isolated Co nanomagnets on a PMN-PT substrate with voltage generated strain, Nano Letters, vol. *17*, 6, pp.3478–3484 (2017)

[27] B. Heinz, T. Brächer, M. Schneider, Q. Wang, B. Lägel, A. M. Friedel, D. Breitbach, S. Steinert, T. Meyer, M. Kewenig, C. Dubs, P. Pirro, A. V. Chumak, Propagation of spin-wave packets in individual nanosized yttrium iron garnet magnonic conduits, Nano Lett., vol. 20, 4220–4227 (2020)

[28] C. O. Avci, A. Quindeau, C.-F. Pai, M. Mann, L. Caretta, A. S. Tang, M. C. Onbasli, C. A Ross, G. S. D. Beach, Current-induced switching in a magnetic insulator, Nature Materials, vol. 16, pp. 309–314 (2017)

[29] J. J. Bauer, E. R. Rosenberg, S. Kundu, K. A. Mkhoyan, P. Quarterman, A. J. Grutter, B. J. Kirby, J. A. Borchers, and C. A. Ross, Dysprosium Iron Garnet Thin Films with Perpendicular Magnetic Anisotropy on Silicon, Adv. Electron. Mater., vol. 6, 1900820 (2020)

[30] M. J. Gross, W. A. Misba, K. Hayashi, D. Bhattacharya, D. B. Gopman, J. Atulasimha, C. A. Ross, Voltage modulated magnetic anisotropy of rare earth iron garnet thin films on a piezoelectric substrate, Appl. Phys. Lett., vol. 121, 252401 (2022)

[31] A. Raja, P.M. M. Gazzali, G. Chandrasekaran, Enhanced electrical and ferrimagnetic properties of bismuth substituted yttrium iron garnets, Physica B: Condensed Matter, vol. 613, 412988 (2021)

[32]. Y. Fan, M. J. Gross, T. Fakhrul, J. Finley, J. T. Hou, S. Ngo, L. Liu and C. A. Ross, Coherent magnon-induced domain-wall motion in a magnetic insulator channel, Nature Nanotechnology, vol. 18, pp. 1000–1004 (2023)

[33] T. Fakhrul, B. Khurana, H. T. Nembach, J. M. Shaw, Y. Fan, G. A. Riley, L. Liu, and C. A. Ross, Substrate-Dependent Anisotropy and Damping in Epitaxial Bismuth Yttrium Iron Garnet Thin Films, Adv. Mater. Interfaces, vol. 10, 2300217 (2023)

[34] L. Soumah, N. Beaulieu, L. Qassym, C. Carrétéro, E. Jacquet, R. Lebourgeois, J. B. Youssef, P. Bortolotti, V. Cros and A. Anane, Ultra-low damping insulating magnetic thin films get perpendicular, Nature Communications, vol. 9, art. no. 3355 (2018)

[35] T. Fakhrul, S. Tazlaru, L. Beran, Y. Zhang, M. Veis, C. A. Ross, Magneto-Optical Bi:YIG Films with High Figure of Merit for Nonreciprocal Photonics, Adv. Optical Mater. 7, 1900056 (2019)

[36] M. Levy, R.M. Osgood, H. Hegde, F.J. Cadieu, R. Wolfe, V.J. Fratello, Integrated optical isolators with sputter-deposited thin-film magnets. IEEE Photon Technol Lett. 8, no.7, pp. 903–905 (1996)

[37] H. Hayashi, S. Iwasa, N.J. Vasa, T. Yoshitake, K. Ueda, S. Yokoyama, S. Higuchi, H. Takeshita, M. Nakahara, Fabrication of Bi-doped YIG optical thin film for electric current sensor by pulsed laser deposition, Applied Surface Science, vol. 197–198, pp. 463-466 (2002)

[38] L. Caretta, S. H. Oh, T. Fakhrul, D. K. Lee, B. H. Lee, S. K. Kim, C. A. Ross, K. J. Lee, G. S. D. Beach, Relativistic kinematics of a magnetic soliton, SCIENCE, vol. 370, 6523, pp. 1438-1442 (2020)

[39] G. G. Siu, C. M. Lee, and Y. Liu, Magnons and acoustic phonons in $Y_{3-x}Bi_xFe_5O_{12}$. Phys. Rev. B 64, 094421 (2001).

[40] Y. C. Wu, K. Garello, W. Kim, M. Gupta, M. Perumkunnil, V. Kateel, S. Couet, R. Carpenter, S. Rao, S. Van Beek, K. K. Vudya Sethu, F. Yasin, D. Crotti, and G. S. Kar, Voltage-Gate Assisted Spin-Orbit Torque Magnetic Random Access Memory for High-Density and Low-Power Embedded Application, Physical Review Applied, vol. 15, 064015 (2021).

[41] M. A. Azam, D. Bhattacharya, D. Querlioz, C. A. Ross, J. Atulasimha, Voltage control of domain walls in magnetic nanowires for energy-efficient neuromorphic devices, Nanotechnology, 31 145201 (2020)

[42] W. A. Misba, M. Lozano, D. Querlioz, J. Atulasimha, Energy Efficient Learning with Low Resolution Stochastic Domain Wall Synapse Based Deep Neural Networks, IEEE Access, 10, 84946 (2022)

[43] K. Hayashi, K. P. Dao, M. J. Gross, L. Ranno, J. X. B. Sia, T. Fakhrul, Q. Du, N. Chatterjee, J. Hu, C. A. Ross, Magneto-Optical Bi-Substituted Yttrium and Terbium Iron Garnets for On-Chip Crystallization via Microheaters, Adv. Optical Mater., vol. 12,2400708 (2024)

[44] N. Jia, Z. Huaiwu, J. Li, Y. Liao, L. Jin, C. Liu, V. G. Harris, Polycrystalline Bi substituted YIG ferrite processed via low temperature sintering, Journal of Alloys and Compounds, vol. 695, pp. 931-936 (2017)

[45] M. J. Gross, J. J. Bauer, S. Ghosh, S. Kundu, K. Hayashi, E. R. Rosenberg, K. A. Mkhoyan, C. A. Ross, Crystallization and stability of rare earth iron garnet/Pt/gadolinium gallium garnet heterostructures on Si, Journal of Magnetism and Magnetic Materials, vol. 564, 170043 (2022)

[46]. T. Wu, P. Zhao, M. Bao, A. Bur, J. L. Hockel, K. Wong, K. P. Mohanchandra, C. S. Lynch, and G. P. Carman, Giant electric-field-induced reversible and permanent magnetization reorientation on magnetoelectric Ni/(011)[Pb (Mg1/3Nb2/3) O3](1− x)–[PbTiO3] x heterostructure, J. Appl. Phys., vol. 109, 124101 (2011).

[47] Y. Lin, L. Jin, H. Zhang, Z. Zhong, Q. Yang, Y. Rao, M. Li, Bi-YIG ferrimagnetic insulator nanometer films with large perpendicular magnetic anisotropy and narrow ferromagnetic resonance linewidth, Journal of Magnetism and Magnetic Materials, vol. 496, 165886 (2020)

[48] P. Hansen, K. Witter, and W. Tolksdorf, Magnetic and magneto-optic properties of lead- and bismuth-substituted yttrium iron garnet films, Physical Review B, vol. 27, 11 (1983)

[49] Y. Zhang, Z. Wang, Y. Wang, C. Luo, J. Li, and D. Viehland, Electric-field induced strain modulation of magnetization in Fe-Ga/Pb(Mg1/3Nb2/3)- PbTiO3 magnetoelectric heterostructures, Journal of Applied Physics, vol. 115, 084101 (2014)

[50] G. Srinivasan, M. I. Bichurin, and J. V. Mantese, Ferromagnetic-ferroelectric layered structures: magnetoelectric interactions and devices, Integrated Ferroelectrics, vol. 71, 45 (2005).

[51] H. Liuyang, P. Freddy, R. Denis, L. Tuami, T. Nicolas, W. Genshui, and P. Philippe, A comparison of converse magnetoelectric coupling effect of YIG film on FE and AFE ceramic substrates, Ferroelectrics, vol. 557, 1 (2020).

[52] D. Pesquera, E. Khestanova, M. Ghidini, S. Zhang, A. P. Rooney, F. Maccherozzi, P. Riego, S. Farokhipoor, J. Kim, X. Moya, M. E. Vickers, N. A. Stelmashenko, S. J. Haigh, S. S. Dhesi, and N. D. Mathur, Large magnetoelectric coupling in multiferroic oxide heterostructures assembled via epitaxial lift-off, Nat. Commun., vol. 11, 3190 (2020).

[53] W. Jahjah, J. P. Jay, Y. le Grand, A. Fessant, A. R. E. Prinsloo, C. J. Sheppard, D. T. Dekadjevi, and D. Spenato, Electrical Manipulation of Magnetic Anisotropy in a Fe81Ga19/Pb (Mg1/3Nb2/3)O3-Pb(Zr$x$Ti1−$x$)O3 Magnetoelectric Multiferroic Composite, Phys. Rev. Appl., vol. 13, 034015 (2020).

[54] S. Fujii, T. Usami, Y. Shiratsuchi, A. M. Kerrigan, A. M. Yatmeidhy, S. Yamada, T. Kanashima, R. Nakatani, V. K. Lazarov, T. Oguchi, Y. Gohda, and K. Hamaya, Strain-induced specific orbital control in a Heusler alloy-based interfacial multiferroics, NPG Asia Mater., vol. 14, 43 (2022).

[55] R. O. Cherifi, V. Ivanovskaya, L. C. Phillips, A. Zobelli, I. C. Infante, E. Jacquet, V. Garcia, S. Fusil, P. R. Briddon, N. Guiblin, A. Mougin, A. A. Unal, F. € Kronast, S. Valencia, B. Dkhil, A. Barthelemy, and M. Bibes, Electric-field control of magnetic order above room temperature, Nat. Mater., vol. 13, 345 (2014).

[56] A. Roe, D. Bhattacharya, J. Atulasimha, Resonant acoustic wave assisted spin-transfer-torque switching of nanomagnets, Appl. Phys. Lett., vol. 115, 112405 (2019)

[57] W. A. Misba, M. M. Rajib, D. Bhattacharya, J. Atulasimha, Acoustic-wave-induced ferromagnetic-resonance-assisted spin-torque switching of perpendicular magnetic tunnel junctions with anisotropy variation, Phys. Rev. Applied, vol. 14, 014088 (2020)

[58] W.-G. Yanga and H. Schmidt, Acoustic control of magnetism toward energy-efficient applications, Appl. Phys. Rev., vol. 8, 021304 (2021)

[59] K. Roy, S. Bandyopadhyay, J. Atulasimha, Hybrid spintronics and straintronics: A magnetic technology for ultra low energy computing and signal processing, Applied Physics Letters, vol. 99, 063108 (2011)

[60] N. Lei, T. Devolder, G. Agnus, P. Aubert, L. Daniel, J.-V. Kim, W. Zhao, T. Trypiniotis, R. P. Cowburn, C. Chappert, D. Ravelosona and P. Lecoeur, Strain-controlled magnetic domain wall propagation in hybrid piezoelectric/ferromagnetic structures, Nature Communications, vol. 4, art. no. 1378 (2013)

Walid Al Misba [1], Miela Josephine Gross [2,3], Kensuke Hayashi [3,4], Daniel B. Gopman [5], Caroline A. Ross [3], Jayasimha Atulasimha [1,6,7]

[1] Mechanical and Nuclear Engineering, Virginia Commonwealth University, Richmond, VA, USA
[2] Electrical Engineering and Computer Science, Massachusetts Institute of Technology, Cambridge, MA, USA
[3] Department of Materials Science and Engineering, Massachusetts Institute of Technology, Cambridge, MA, USA
[4] Department of Materials Physics, Graduate School of Engineering, Nagoya University, Nagoya, Japan
[5] Materials Science & Engineering Division, National Institute of Standards and Technology, Gaithersburg, MD, USA
[6] Electrical and Computer Engineering, Virginia Commonwealth University, Richmond, VA, USA
[7] Department of Physics, Virginia Commonwealth University, Richmond, VA, USA

**S1. Domain reversal Study using Magneto Optic Kerr Effect (MOKE) microscopy for In-Plane direction $\hat{y}$ // $[01\bar{1}]$:**

The sample is first poled by applying 450 V and subsequently relaxed to 0 V. The voltage is kept at 0 V and the sample is saturated by applying an external field of -70 mT along the in-plane direction $\hat{y}$. Domains with black contrast are prominent and occupy the observed regions as can be seen in Fig. S1a. The reversal field is then set to +27 mT and MOKE images are acquired by increasing the voltages in steps of 50 V up to 450 V. With the application of voltage, white contrast reserved domains are nucleated across the observed regions and the number of reversed domains increases with voltage. In the poled sample, at 0 V, $\hat{y}$-axis is the easy axis of magnetization due to the remnant compressive strain from the PMN-PT substrate [S1, S2]. This is evident from the square-shaped hysteresis loops as in Fig. 4c (in the main manuscript) and the prominent domain wall nucleation and growth in Fig. 4d at 0 V. Due to the higher magnetic anisotropy and coercive field (30 mT $\pm$ 2 mT), fewer numbers of reversed domains are seen for +27 mT field at 0 V. As we increase the voltage, the $\hat{y}$-direction becomes harder for the magnetization due to the tensile strain from PMN-PT, thus the coercive field decreases. Correspondingly, the quantity of reversed domains increases. Significant changes in the quantity of nucleated domains are not observed beyond a bias of 400 V or higher across the PMN-PT substrate.

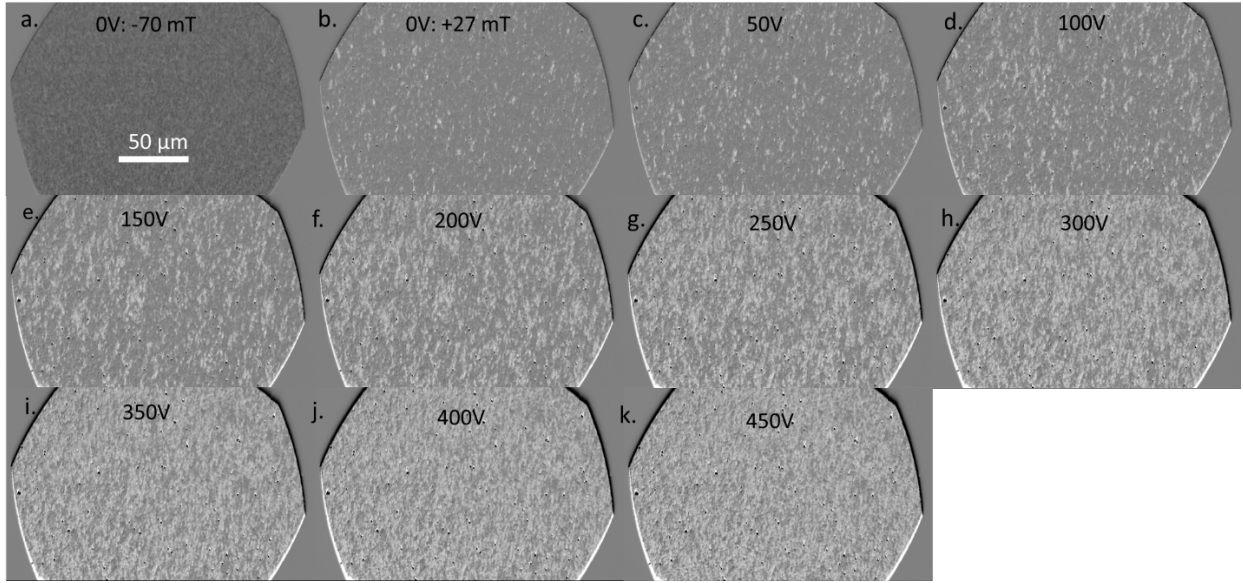

**Fig. S1:** MOKE images showing the saturated domains and reversal of the domains for varying amplitude voltages at a fixed reversal field along the in-plane direction $\hat{y}$ // $[01\bar{1}]$. a. The sample is poled at 450 V and relaxed to 0 V and then saturated with -70 mT field. The external field is then fixed at + 27 mT, while the voltage remains at b. 0 V and increased to c. 50 V d. 100 V e. 150 V f. 200 V g. 250 V h. 300 V i. 350 V j. 400 V and k. 450 V.

**S2: Ferromagnetic Resonance:**

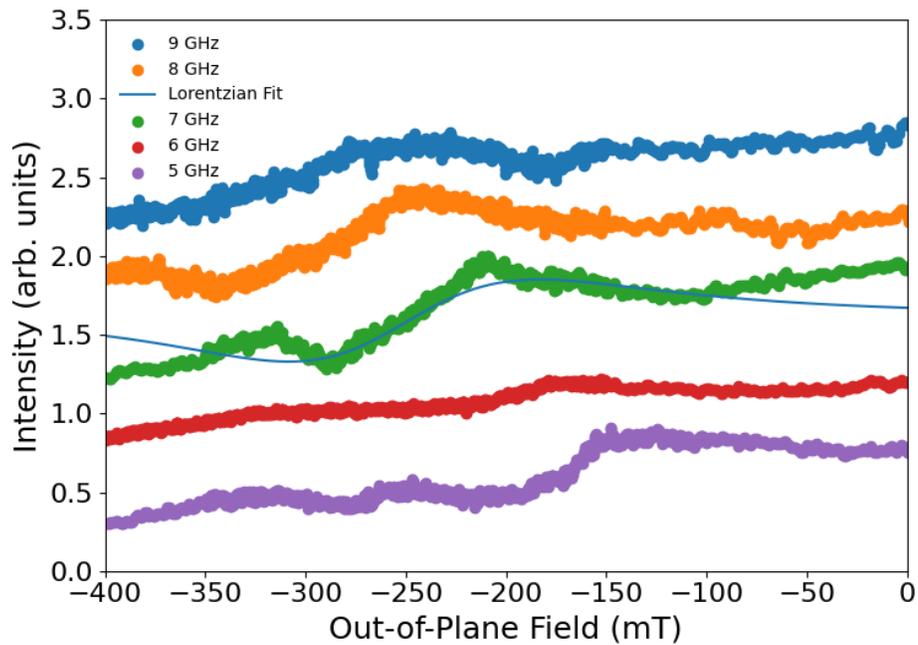

**Fig. S2**: FMR data for fused silica/ Bi-YIG measured at frequencies of 5 – 9 GHz, with an example of a Lorentzian fit.

References:

[S1]. T. Wu, P. Zhao, M. Bao, A. Bur, J. L. Hockel, K. Wong, K. P. Mohanchandra, C. S. Lynch, and G. P. Carman, Giant electric-field-induced reversible and permanent magnetization reorientation on magnetoelectric Ni/(011)[Pb (Mg1/3Nb2/3) O3](1− x)–[PbTiO3] x heterostructure, J. Appl. Phys. 109, 124101 (2011).

[S2]. M. J. Gross, W. A. Misba, K. Hayashi, D. Bhattacharya, D. B. Gopman, J. Atulasimha, C. A. Ross, Voltage modulated magnetic anisotropy of rare earth iron garnet thin films on a piezoelectric substrate, Appl. Phys. Lett. 121, 252401 (2022)